\documentclass[11pt,twoside]{article}

\usepackage{asp2006}
\usepackage{epsf}
\usepackage{psfig}
\usepackage{lscape}

\newcommand{\HI}{\hbox{{\rm H}\kern 0.1em{\sc i}}}
\newcommand{\CaII}{\hbox{{\rm Ca}\kern 0.1em{\sc ii}}}

\begin{document}
\title{Searching for the Precursors of Life in External Galaxies}   
\author{B. Lawton \& C. W. Churchill}   
\affil{Department of Astronomy, MSC 4500, New Mexico State University, P.O. Box 30001, Las Cruces, NM 88003}    

\author{B. A. York \& S. L. Ellison}
\affil{Department of Physics and Astronomy, University of Victoria, P.O. Box 3055, Station CSC, Victoria, BC V8W 3P6, Canada}

\author{T. P. Snow}
\affil{Center for Astrophysics and Space
Astronomy, University of Colorado at Boulder, 389 UCB, Boulder, CO 80309}

\author{R. A. Johnson}
\affil{Oxford Astrophysics, Denys Wilkinson Building, Keble Road, Oxford OX1 3RH, UK; raj@astro.ox.ac.uk}

\author{S. G. Ryan}
\affil{Centre for Astrophysics Research, University of Hertfordshire, College Lane, Hatfield AL10 9AB, UK}

\author{C. R. Benn}
\affil{Isaac Newton Group, Apartado 321, E-38700 Santa Cruz de La Palma, Spain}

\begin{abstract} 

Are the organic molecules crucial for life on Earth abundant in
early-epoch galaxies?  To address this, we searched for organic
molecules in extragalactic sources via their absorption features,
known as diffuse interstellar bands (DIBs).  There is strong evidence
that DIBs are associated with polycyclic aromatic hydrocarbons (PAHs)
and carbon chains.  Galaxies with a preponderance of DIBs may be the
most likely places in which to expect life.

We use the method of quasar absorption lines to probe intervening
early-epoch galaxies for the DIBs.  We present the equivalent width
measurements of DIBs in one neutral hydrogen (\HI) abundant galaxy and
limits for five DIB bands in six other {\HI}-rich galaxies (damped
Lyman-$\alpha$ systems--DLAs).  Our results reveal that {\HI}-rich
galaxies are dust poor and have significantly lower reddening than
known DIB-rich Milky Way environments.  We find that DIBs in
{\HI}-rich galaxies do not show the same correlation with hydrogen
abundance as observed in the Milky Way; the extragalactic DIBs are
underabundant by as much as 10 times.  The lower limit gas-to-dust
ratios of four of the {\HI}-rich early epoch galaxies are much higher
than the gas-to-dust ratios found in the Milky Way.  Our results
suggest that the organic molecules responsible for the DIBs are
underabundant in {\HI}-rich early epoch galaxies relative to the Milky
Way.

\end{abstract}


\section{Introduction}   

Since their discovery in 1921 \citep{hege22}, the diffuse intersteller
bands (DIBs) have remained the longest known interstellar absorption
features without a positive identification.  There have been several
hundred DIBs discovered to date \citep{jenn94,tuai00,wese00}.  The
DIBs span the visible spectrum between 4000 and 13000~{\AA}.  Despite
no positive identifications, several likely organic molecular
candidates have emerged as the sources of the DIBs, including
polycyclic aromatic hydrocarbons (PAHs), fullerenes, long carbon
chains, and polycyclic aromatic nitrogen heterocycles (PANHs)
\citep{herb95, snow01, cox06a, hudg05}.  The organic-molecular origin
of the DIBs may give them an importance to astrobiology; they are now
considered an important early constituent to the inventory of organic
compounds on Earth \citep{bada02}.

There are several environmental factors that are known to enhance or
inhibit DIB strengths.  Thus, measuring DIB strengths can give clues
to the environments of galaxies like those with high neutral hydrogen
(\HI) content, known as damped Lyman-$\alpha$ systems (DLAs).
Measuring DIB strengths allow researchers to explore the quantities of
gas, metallicity, dust, and radiation in galaxies and their influence
on the strengths of the bands.  Furthermore, observing DIBs in
galaxies at a higher redshift, $z$, (distance or lookback time related
to the expansion of the Universe) allows us to test the evolution of
these organics in cosmic time.

Two important environmental factors that are often probed in galaxies
are the {\HI} content and the reddening.  The {\HI} content in DLAs is
typically measured as a column density via their Lyman-$\alpha$ line
in absorption, as observed using a bright background source such as a
quasar.  The column density is the number of atoms of a certain
element as seen along a line-of-sight projected to a unit area at the
observer (typically in units of atoms cm$^{-2}$ and denoted as
$N$(\HI) for neutral hydrogen).  The reddening is a measure of the
dust content in the galactic environment.  Reddening is expressed as
$E(B-V)$, or the magnitude of blue light minus visible light observed
relative to what is expected from typical stars or galaxies.  Because
dust preferentially obscures blue light, a high reddening is a
signature for a high dust content.  $N$(\HI) is a measure of the gas
phase and $E(B-V)$ is a measure of the dust phase of the ISM in
galaxies.

Due to their relatively weak absorption strengths, the DIBs have been
difficult to detect in extragalactic sources.  Aside from the hundreds
of detections within the Milky Way \citep{jenn94,tuai00, wese00}, DIBs
have been detected in the Magellanic Clouds
\citep{welt06,cox06b,cox07}, seven starburst galaxies \citep{heck00},
the active galaxy Centaurus A via supernova 1986A \citep{rich87},
spiral galaxy NGC 1448 via Supernovae 2001el and 2003hn
\citep{soll05}, one DLA galaxy at $z=0.524$ toward the quasar
AO~0235+164 \citep{junk04, lawt06}, and one galaxy selected by singly
ionized calcium (\CaII), J0013--0024, at $z=0.157$ from the Sloan
Digital Sky Survey \citep{elli07}.

We further the knowledge of DIB lines in extragalactic environments by
cataloguing the strengths of the $\lambda4428$, $\lambda5780$,
$\lambda5797$, $\lambda6284$, and $\lambda6613$ DIBs relative to the
$E(B-V)$ and $N$(\HI) content of each of the seven DLAs in our sample.
Observations were obtained, with seven facilities, of seven DLAs
toward six QSO sightlines.  The facilities and instruments used for
this project are the VLT/FORS2, VLT/UVES, APO/DIS, Keck/HIRES,
WHT/ISIS, and Gemini/GMOS-S.

\section{Analysis \& Results}

There are two detections included in this work, the $\lambda$5705 and
$\lambda$5780 DIBs first reported by \citet{lawt06}, in the $z=0.524$
DLA toward AO~0235+164.  For all other DLAs in our sample, we report
upper limits on the $\lambda$4428, $\lambda$5780, $\lambda$5797,
$\lambda$6284, and $\lambda$6613 DIB equivalent widths. We measured
the equivalent width limits using a generalized method of the
\citet{schn93} technique for finding lines and limits.  We compare our
measured limits to the expected DIB equivalent widths from the known
Milky Way DIB--$E(B-V)$ and DIB--$N$(\HI) relations \citep{welt06}.
The $N$(\HI) quantities are known for the DLAs; however,
\citet{junk04} published the only reddening known for the DLA galaxies
in our sample, AO~0235+164, with a measured $E(B-V)=0.23$.  We
estimate the upper limit to the reddening using our equivalent width
limits and the Milky Way DIB--$E(B-V)$ correlation.  Our equivalent
width limits are robust enough to constrain the upper reddening limits
near the $E(B-V)<0.04$ limit found by \citet{elli05} for the highest
redshift DLA galaxies.

The results from the $N$(\HI) model suggests that the organics that
give rise to the DIBs in DLAs are underabundant relative to Milky Way
sightlines of the same hydrogen column density.  Fig.~1 shows this by
plotting the measured equivalent widths and upper equivalent width
limits for the DLAs in our sample.  The line is the best-fit to the
Milky Way data from \citet{welt06}.  The Milky Way points are observed
to lie within the dotted region while the Large Magellanic Cloud
sightlines are observed to lie within the dashed region.  The Small
Magellanic Cloud sightlines are all within the dot-dashed region.  The
$\lambda$6284 DIB gives the best constraints and shows that this DIB
is at least 4-10 times weaker in four of our DLAs compared to what is
expected in the Milky Way.  As is the case for the Magellanic Clouds,
the Milky Way DIB--$N$(\HI) relation does not apply to DIBs in DLAs.
Many environmental factors can potentially work to inhibit the
organics responsible for the DIBs so this alone can not be used as
evidence that the environments of DLAs are similar to the Large or
Small Magellanic Cloud.  For example, the dust content, ionizing
radiation, and metallicity may be quite different in DLAs relative to
the Magellanic Clouds.

\begin{figure}[!ht]
\plotfiddle{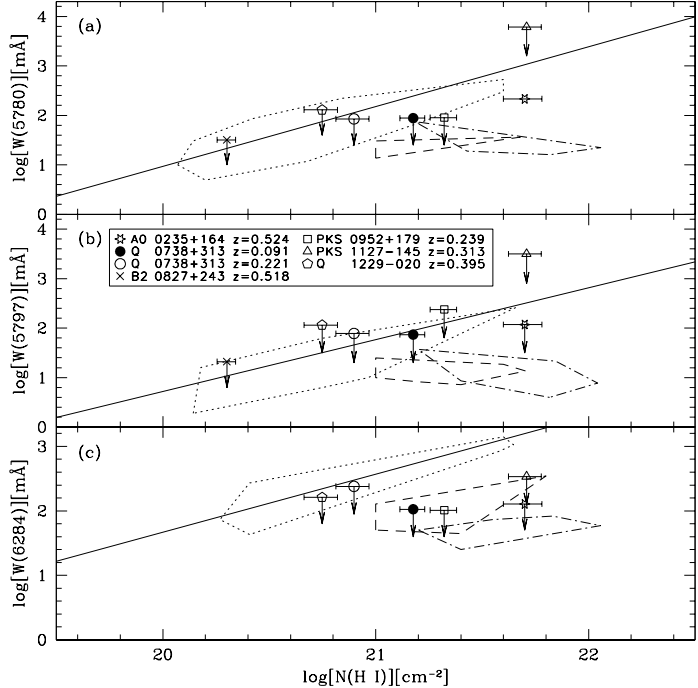}{3.4in}{0}{50.}{50.}{-160}{-90}
\caption{The DIB equivalent width--$N$(\HI) relations \citep{welt06} with our DLAs added.  ---($a$) $\lambda$5780 DIB.  ---($b$) $\lambda$5797 DIB.  ---($c$) $\lambda$6284 DIB.  The solid lines are the best-fit weighted Milky Way lines.  The region enclosed by the dotted lines contain the Milky Way data.  The regions enclosed by the dashed lines contain the LMC data.  The regions enclosed by the dot-dash lines contain the SMC data.  Error bars are 1~$\sigma$, and upper limits are marked with arrows.  The vertical error bars for AO~0235+164 in panel ($a$) are smaller than the point size and all values for this DLA are from \citet{lawt06}.}
\end{figure}

\begin{figure}[!ht]
\plotfiddle{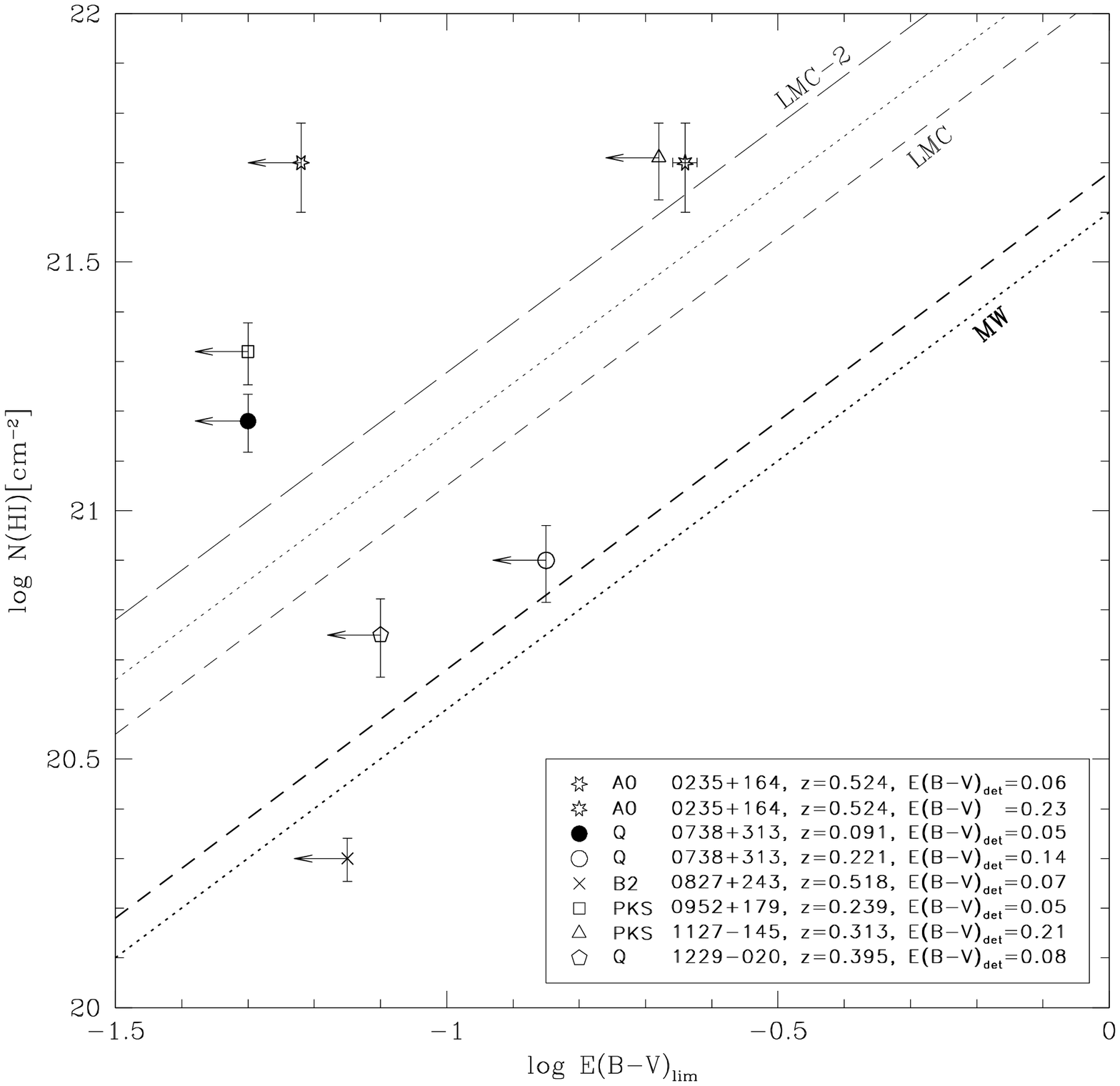}{2.2in}{0}{50.}{32.}{-160}{-60}
\caption{The gas-to-dust ratios of the DLAs in our sample relative to measured values in the Milky Way (MW) and the Large Magellanic Cloud (LMC).  The figure is modified from \citet{cox06b}.  The plot measures the log column density [cm$^{-2}$] versus the upper limit to the log reddening for each DLA determined.  The A0~0235+164 reddening measurement of 0.23$\pm$0.01 is from \citet{junk04}.  The top three lines represent the LMC while the bottom two lines represent the MW.  The long-dashed LMC line gives the gas-to-dust ratio of 19.2~$\times$~10$^{21}$~cm$^{-2}$~mag$^{-1}$ from the LMC-2 data of \citet{gord03}.  The dotted LMC line is a linear fit to the LMC data in \citet{cox06b} and gives a gas-to-dust ratio of 14.3~$\times$~10$^{21}$~cm$^{-2}$~mag$^{-1}$.  The short-dashed line is the average LMC regions from \citep{gord03} and has a gas-to-dust ratio of 11.1~$\times$~10$^{21}$~cm$^{-2}$~mag$^{-1}$.  The dashed MW line gives a gas-to-dust ratio of 4.8~$\times$~10$^{21}$~cm$^{-2}$~mag$^{-1}$ \citep{bohl78}, and the dotted MW line is the fit to the Milky Way data from \citet{cox06b} which yields a ratio of 4.03~$\times$~10$^{21}$~cm$^{-2}$~mag$^{-1}$.  Several of the DLAs in our sample are consistent with having higher gas-to-dust ratios than the MW and the LMC.}
\end{figure}

From our equivalent width limits we can estimate the upper limit to
the reddening assuming the Milky Way DIB--$E(B-V)$ relation holds for
DLAs.  There are little data on DIBs in DLAs; however, \citet{elli07}
create a fit to all known extragalactic points for the $\lambda$5780
DIB.  The $\lambda$5780 DIB--$E(B-V)$ relation appears to remain valid
when the extragalactic equivalent width measurements are included,
although the slope is slightly steeper.  Our upper limits for $E(B-V)$
yield lower limits to the gas-to-dust ratios for our DLAs; these
results are shown in Fig.~2.  $E(B-V)_{lim}$ are the upper limits for
the reddening determined by our best equivalent width limits.  The
best-fit lines for the Milky Way and the Large Magellanic Cloud are
given from the literature.  Our limits are robust enough to constrain
the gas-to-dust ratio for the DLAs as being much higher than the
Milky Way for four of the DLAs.  The discrepency between the limit and
the measurement of the DLA toward AO~0235+164 may be due to
problematic measurements.  Our best gas-to-dust lower limit comes from
the reddening constraints for the $\lambda$6284 DIB; however, there is
a large atmospheric absorption band at the DIB's expected location.
It could also be that the $\lambda$6284 DIB--$E(B-V)$ relation is
different for DLAs then is measured for the Milky Way.  The reddening
limits allow us to extend the results by \citet{elli05} to lower
redshift DLAs.  The $E(B-V)<0.04$ from \citet{elli05} appears to also
be a robust result for moderate redshift DLAs.

\section{Implications for Astrobiology}

Despite having comparable {\HI} abundances to the Milky Way, DLA
galaxies have DIB strengths that are weaker than expected.  The
weakness of DIBs in DLA selected galaxies suggests that the
environments of these early epoch {\HI}-rich galaxies are much less
suitable to create and/or sustain organic molecules.  DLA galaxies do
have considerably lower reddening than the Milky Way regions where
DIBs are observed.  Our results imply that the low reddening in DLAs
will inhibit the DIBs if they are present.  Reddening is dependent on
dust grain sizes and abundances.  It is conceivable that the organics
require dust grains for their formation or are created from the same
sources, such as carbon stars \citep{herb95}.  A large carbon
abundance may be required for the DIB--$E(B-V)$ relations to hold
true.  Therefore, it is feasible that early epoch {\HI} galaxies are
not conducive to the formation of the organics because they lack the
necessary carbon.  However, \citet{herb95} points out that the scatter
in the Milky Way DIB--$E(B-V)$ relation is real, ie not due to noise
or systematic errors.  Thus, other environmental factors such as the
local ionizing radiation field and the metallicity may be important
\citep{cox07}.  Little is published about the radiation in the DLA
galaxies in our sample.  Their metallicities are known to be low, on
the order of $\sim$0.1 Solar metallicity, with the exception of
AO~0235+164, which is metal abundant \citep{junk04}.  Based on
published work, it is not surprising that the DLA galaxy in our sample
with the highest reddening and metallicity has the only detected DIBs.
Thus, the lack of dust and metals plays a large role in inhibiting the
organics in our sample of early epoch {\HI}-rich galaxies; whereas,
the abundance of {\HI} is not a strong determinent in their strengths.
Our DLA selection method is not representative of all early epoch
galaxies; it is biased toward galaxies with low reddening.  Another
selection method for early epoch galaxies may be more fruitful.

\acknowledgements 

BL acknowledges the support of NASA via the Graduate Student
Researchers Program (GSRP).  BL also thanks the NASA Astrobiology
Institute and the National Science Foundation for travel support
arranged through the conference organizers.



\begin{thebibliography}{}

\bibitem[Bada \& Lazcano(2002)]{bada02} Bada, J. L. \& Lazcano, A. 2002, Science, 296, 1982
\bibitem[Bohlin et al.(1978)]{bohl78} Bohlin, R. C., Savage, B. D., \& Drake, J. F. 1978, \apj, 224, 132
\bibitem[Cox \& Spaans(2006a)]{cox06a} Cox, N. L. J., \& Spaans, M. 2006a, \aap, 451, 973
\bibitem[Cox et al.(2006b)]{cox06b} Cox, N. L. J., Cordiner, M. A., Cami, J., Foing, B. H., Sarre, P. J., Kaper, L., \& Ehrenfreund, P. 2006b, \aap, 447, 991
\bibitem[Cox et al.(2007)]{cox07} Cox, N. L. J., Cordiner, M. A., Ehrenfreund, P., Kaper, L., Sarre, P. J., Foing, B. H., Spaans, M., Cami, J., Sofia, U. J., Clayton, G. C., Gordon, K. D., \& Salama, F. 2007, \aap, 470, 941
\bibitem[Ellison et al.(2005)]{elli05} Ellison, S. L., Hall, P. B., \& Lira, P. 2005, \aj, 130, 1345
\bibitem[Ellison et al.(2007)]{elli07} Ellison, S. L., York, B. A., Murphy, M. T., Zych, B. J., Smith, A. M., \& Sarre, P. J. 2007, preprint (arXive:0710.0901)
\bibitem[Gordon et al.(2003)]{gord03} Gordon, K. D., Clayton, G. C., Misselt, K. A., Landolt, A. U., \& Wolff, M. J. 2003, \apj, 594, 279
\bibitem[Heckman \& Lehnert(2000)]{heck00} Heckman, T. M. \& Lehnert, M. D. 2000, \apj, 537, 690
\bibitem[Heger(1922)]{hege22} Heger, M. L. 1922, Lick Observatory Bull. 10, 337, 146
\bibitem[Herbig(1995)]{herb95} Herbig, G. H. 1995, \araa, 33, 19
\bibitem[Hudgins et al.(2005)]{hudg05} Hudgins, D. M., Bauschlicher, C. W., Jr., \& Allamandola, L. J. 2005, \apj, 632, 316
\bibitem[Jenniskens et al.(1994)]{jenn94} Jenniskens, P., \& Desert, F. -X. 1994, \aap, 106, 39
\bibitem[Junkkarinen et al.(2004)]{junk04} Junkkarinen, V. T., Cohen, R. D., Beaver, E. A., Burbidge, E. M., \& Lyons, R. W. 2004, \apj, 614, 658
\bibitem[Rich(1987)]{rich87} Rich, R. M. 1987, \aj, 94, 651
\bibitem[Schneider et al.(1993)]{schn93} Schneider, D. P., Hartig, G. F., Jannuzi, B. T., et al. 1993, \apjs, 87, 45
\bibitem[Snow(2001)]{snow01} Snow, T. P. 2001, Spectrochimica Acta Part A, 57, 615
\bibitem[Sollerman et al.(2005)]{soll05} Sollerman, J., Cox, N., Mattila, S., Ehrenfreund, P., Kaper, L., Leibundgut, B., \& Lundqvist, P. 2005, \aap, 429, 559
\bibitem[Tuairisg et al.(2000)]{tuai00} Tuairisg, S. O., Cami, J., Foing, B. H., Sonnentrucker, P., \& Ehrenfreund, P. 2000, \aap, 142, 225
\bibitem[Welty et al.(2006)]{welt06} Welty, D. E., Federman, S. R., Gredel, R., Thorburn, J. A., \& Lambert, D. L. 2006, \apjs, 165, 138
\bibitem[Weselak et al.(2000)]{wese00} Weselak, T., Schmidt, M., \& Krelowski, J. 2000, \aap, 142, 239
\bibitem[York et al.(2006)]{lawt06} York, B. A., Ellison, S. L., Lawton, B., Churchill, C. W., Snow, T. P., Johnson, R. A., \& Ryan, S. G. 2006, \apj, 647, L29
\end{thebibliography}
\end{document}